\documentclass[twocolumn,floats,prl,aps,unsortedaddress,superscriptaddress]{revtex4-1}

\usepackage[dvips]{graphicx}
\usepackage{amsfonts}
\usepackage{amsmath}
\usepackage{amssymb}
\usepackage{exscale}
\usepackage{color}	
\usepackage{braket}	
\usepackage{multirow}	
\usepackage{epstopdf}

\usepackage[normalem]{ulem}    

\begin{document}
\title{Orientational tuning of the Fermi sea of confined electrons 
at the SrTiO$_3$ $(110)$ and $(111)$ surfaces}

\author{T.~C.~R\"odel}
\affiliation{CSNSM, Universit\'e Paris-Sud and CNRS/IN2P3, 
            B\^atiments 104 et 108, 91405 Orsay cedex, France}
\affiliation{Synchrotron SOLEIL, L'Orme des Merisiers, 
            Saint-Aubin-BP48, 91192 Gif-sur-Yvette, France}
\author{C.~Bareille}
\affiliation{CSNSM, Universit\'e Paris-Sud and CNRS/IN2P3,
            B\^atiments 104 et 108, 91405 Orsay cedex, France}
\author{F.~Fortuna}
\affiliation{CSNSM, Universit\'e Paris-Sud and CNRS/IN2P3, 
            B\^atiments 104 et 108, 91405 Orsay cedex, France}
\author{C.~Baumier}
\affiliation{CSNSM, Universit\'e Paris-Sud and CNRS/IN2P3,
            B\^atiments 104 et 108, 91405 Orsay cedex, France}
\author{F.~Bertran}
\affiliation{Synchrotron SOLEIL, L'Orme des Merisiers, 
            Saint-Aubin-BP48, 91192 Gif-sur-Yvette, France}
\author{P.~Le~F\`evre}
\affiliation{Synchrotron SOLEIL, L'Orme des Merisiers, 
            Saint-Aubin-BP48, 91192 Gif-sur-Yvette, France}
\author{M.~Gabay}
\affiliation{Laboratoire de Physique des Solides, Universit\'e Paris-Sud and CNRS, 
            B\^atiment 510, 91405 Orsay, France}
\author{O.~Hijano~Cubelos}
\affiliation{Laboratoire de Physique des Solides, Universit\'e Paris-Sud and CNRS, 
            B\^atiment 510, 91405 Orsay, France}
\author{M.~J.~Rozenberg}
\affiliation{Laboratoire de Physique des Solides, Universit\'e Paris-Sud and CNRS, 
            B\^atiment 510, 91405 Orsay, France}
\affiliation{Depto. de F\'isica - IFIBA Conicet, FCEN, UBA, Ciudad Universitaria P.1, 
            1428, Buenos Aires, Argentina}
\author{T.~Maroutian}
\affiliation{Institut d'Electronique Fondamentale, Universit\'e Paris-Sud and CNRS, 
            B\^atiment 220, 91405 Orsay, France}
\author{P.~Lecoeur}
\affiliation{Institut d'Electronique Fondamentale, Universit\'e Paris-Sud and CNRS, 
            B\^atiment 220, 91405 Orsay, France}
\author{A.~F.~Santander-Syro}
\email{andres.santander@csnsm.in2p3.fr}
\affiliation{CSNSM, Universit\'e Paris-Sud and CNRS/IN2P3, 
            B\^atiments 104 et 108, 91405 Orsay cedex, France}

\begin{abstract}
	We report the existence of confined electronic states at the $(110)$ and $(111)$  
	surfaces of SrTiO$_3$.
    Using angle-resolved photoemission spectroscopy, 
	we find that the corresponding Fermi surfaces,
	subband masses, and orbital ordering 
	are different from the ones at the $(001)$ surface of SrTiO$_3$.  
	This occurs because the crystallographic symmetries of the surface 
	and sub-surface planes, and the electron effective masses along the confinement direction, 
	influence the symmetry of the electronic structure and the orbital ordering 
	of the $t_{2g}$ manifold.
	Remarkably, our analysis of the data also reveals that the carrier concentration 
	and thickness are similar for all three surface orientations, 
	despite their different polarities.
	The orientational tuning of the microscopic properties of two-dimensional electron states
	at the surface of SrTiO$_3$ echoes the tailoring of macroscopic 
	(\emph{e.g.} transport) properties reported recently 
	in LaAlO$_3$/SrTiO$_3$ $(110)$ and $(111)$ interfaces, 
	and is promising for searching new types of 2D electronic states 
	in correlated-electron oxides.
\end{abstract}
\maketitle

Two-dimensional electron gases (2DEGs) in transition-metal oxides (TMOs)
present remarkable phenomena that make them unique from a fundamental viewpoint
and promising for applications~\cite{Takagi2010,Mannhart2010}.
For instance, heterostructures grown on the $(001)$ surface of SrTiO$_3$,
a TMO insulator with a large band-gap of $\sim 3.5$~eV,
can develop 2DEGs showing metal-to-insulator transitions~\cite{Thiel2006}, 
superconductivity~\cite{Caviglia2008}, 
or magnetism~\cite{Brinkman2007,Salluzzo2013}. 
Recently, 2DEGs at the $(111)$ and $(110)$ interfaces of LaAlO$_3$/SrTiO$_3$ 
were also reported~\cite{Herranz2012}.
The latter showed a highly anisotropic conductivity~\cite{Annadi2013} 
and a superconducting state spatially more extended 
than the one at the $(001)$ interface~\cite{Herranz2013}.
Interestingly, theoretical works have also predicted that exotic, possibly topological, 
electronic states might occur at interfaces composed of $(111)$ bilayers 
of cubic TMOs~\cite{Xiao2011,Yang2011,Ruegg2011,Doennig2013},
as two $(111)$ planes of transition-metal ions form a honeycomb lattice, 
similar to the one found in graphene.
In this context, the discoveries that 2DEGs can also be created at the bare $(001)$ surfaces 
of SrTiO$_3$~\cite{Santander-Syro2011,Meevasana2011,Plumb2013}
and KTaO$_3$~\cite{King2012,Santander-Syro2012},
and more recently at the $(111)$ surface of KTaO$_3$~\cite{Bareille2014},
opened new roads in the fabrication and study of different types of 2DEGs in TMOs
--in particular using surface-sensitive spectroscopic techniques,
which give direct information about the Fermi surface and subband structure
of the confined states. 
The origin of the confinement is attributed to a local doping 
of the surface region due to oxygen vacancies and/or lattice distortions.

Here we show that new types of 2DEGs can be directly tailored 
at the bare $(110)$ and $(111)$ surfaces of SrTiO$_3$.
Imaging their electronic structure \emph{via} angle-resolved photoemission spectroscopy (ARPES), 
we find that their Fermi surfaces, subband masses, and orbital ordering 
are different from the ones of the 2DEG 
at the SrTiO$_3$~$(001)$ surface~\cite{Santander-Syro2011,Meevasana2011}
and the ones predicted for the bulk,
being thus uniquely sensitive to the confining crystallographic direction. 
This occurs because the crystallographic symmetries of the 2DEG plane, 
and the electron effective masses along the confinement direction, 
influence the symmetry of the electronic structure and the orbital ordering 
of the $t_{2g}$ orbitals.
Furthermore, the observed carrier concentrations and 2DEG thicknesses 
for different surfaces allow us to showcase the impact of oxygen  vacancies 
and of the polar discontinuity on distinctive features of the confined conducting sheet.

\begin{figure*}
  \begin{center}
   	  \includegraphics[clip, width=14cm]{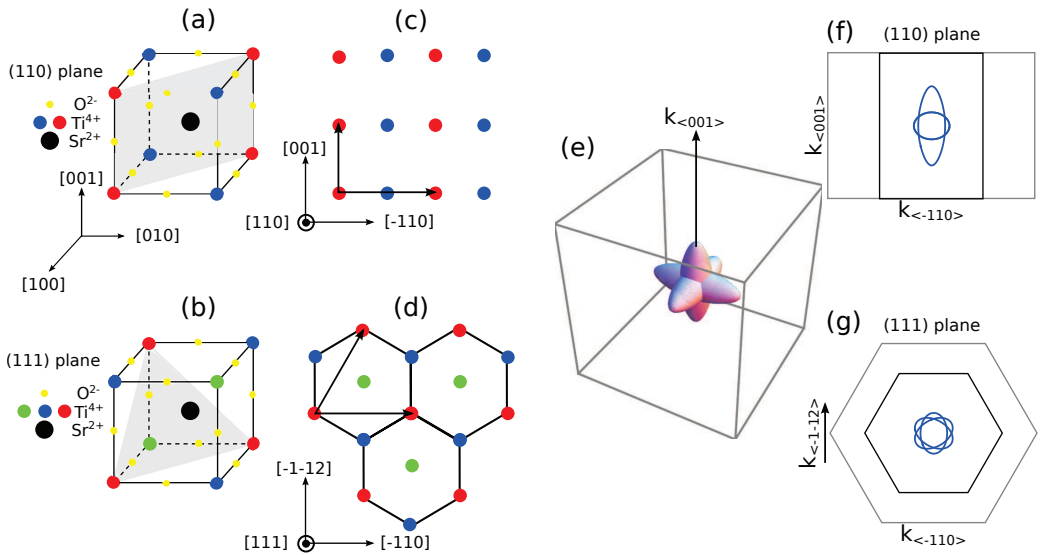}
  \end{center}
  \caption{\label{fig:fig1} \footnotesize{
  		(a,~b) Unit cell of the cubic perovskite lattice of SrTiO$_3$.
        The grey planes are the $(110)$ and $(111)$ planes, respectively. 
        The yellow dots represent the O$^{2-}$ anions, the black dot in the center 
  		the Sr$^{2+}$ cation, and the red/green/blue dots the Ti$^{4+}$ cations  
  		in different $(110)$ or $(111)$ planes. 		
  		Both orientations are highly polar, 
  		as the crystal is built of alternating layers of (SrTiO)$^{4+}$ and 
  		(O$_2)^{4-}$ or Ti$^{4+}$ and (SrO$_3)^{4-}$. 
  		(c,~d) Ti$^{4+}$ cations of the crystal lattice at the $(110)$ and $(111)$ planes. 
  		The black arrows indicate the lattice vectors of the Ti$^{4+}$ cations 
  		in one $(110)$ or $(111)$ plane. 
  		As indicated by the black lines in panel~(d), 
  		a $(111)$-bilayer of Ti$^{4+}$ cations forms a honeycomb lattice. 
  		(e) Bulk Fermi surface, calculated using a tight-binding model 
  		with an unrealistically large value of $10^{21}$~cm$^{-3}$ 
  		for the bulk carrier density, intended to make the Fermi surface visible.
  		Such carrier density is at least \emph{three orders of magnitude} higher 
  		than the bulk carrier density of the samples prepared for this study. 		
  		(f,~g) Cross section of the bulk Fermi surface in (e) 
  		along the $(110)$ and $(111)$ planes, respectively.
  		The grey lines show the cross section of the bulk 3D Brillouin zone 
  		through a $\Gamma$ point, while the black lines correspond 
  		to the surface Brillouin zone.
  		}
  	} 
\end{figure*}
The confined states were either created by fracturing the samples in vacuum 
or by chemically and thermally preparing the surfaces \emph{in situ},
and studied through ARPES at the Synchrotron Radiation Center 
(SRC, University of Wisconsin, Madison)
and the Synchrotron Soleil (France). 
The sample preparation, similar to the one in references~\cite{Biswas2011,Chang2008}, 
is detailed in the Supplemental Material~\cite{Supplement}.  
All through this paper, we describe the crystal structure in a cubic basis of unit-cell vectors, 
and note as $[hkl]$ the crystallographic directions in real space, 
$\langle hkl \rangle$ the corresponding  directions in reciprocal space,
and as $(hkl)$ the planes orthogonal to those directions.

The major difference between the confined states 
at various surface orientations of SrTiO$_3$ 
originates from the different symmetries of the corresponding crystal planes: 
4-fold for the $(001)$ plane, 2-fold for the $(110)$ surface, and 6-fold for the $(111)$ surface.
Another difference is the polar character of the surface. 
Thus, while the $(001)$ terminations, namely SrO or TiO$_2$, are nominally non-polar,
the $(110)$ terminations are alternatively $($SrTiO$)^{4+}$ and $($O$_2)^{4-}$, 
and the $(111)$ terminations are either Ti$^{4+}$ or $($SrO$_3)^{4-}$.
These different surface symmetries and their polarity 
are illustrated in figures~\ref{fig:fig1}(a-d).
Note in particular, from figure~\ref{fig:fig1}(d), that
a $(111)$-type bilayer of Ti$^{4+}$ cations forms a honeycomb lattice,
as noted in Ref.~\cite{Xiao2011}.

\begin{table}[b]
 \caption{\label{Table:Masses} \footnotesize{
 			Effective light (L) and heavy (H) masses predicted by a TB model 
			in the bulk (first row) and experimental in-plane masses of the 2DEGs 
 			at the $(001)$, $(110)$, and $(111)$ surfaces (other rows)
 			along the different high-symmetry directions of the crystal lattice (columns) of SrTiO$_{3}$. 
 			In the bulk, all the effective masses along $\langle 111 \rangle$ are identical.
 			} 
  		 }
 \begin{center}
  \begin{tabular}{c | c c | c c | c c | c}
    \hline \hline
        &  \multicolumn{2}{|c|}{$m_{100}/m_e$}  & \multicolumn{2}{|c|}{$m_{110}/m_e$}
        &  \multicolumn{2}{|c|}{$m_{11\bar{2}}/m_e$}  & $m_{111}/m_e$ \\
  	\hline
  	    & $L$ & $H$ & $L$ & $H$ & $L$ & $H$ &   \\
  	\hline
	Theory bulk$^{a}$  & $1.06$ & $7.16$	& $1.06$ & $1.85$ & $1.24$ & $2.46$ & $1.48$ \\
	\hline
	  SrTiO$_{3}$$(001)$ & $0.7^{b}$ & $10.0^{b}$	 & $0.7^{c}$ & $1.3^{c}$ 
  	           & $0.8^{c}$ & $1.8^{c}$ & $1.0^{c}$ \\
    SrTiO$_{3}$$(110)$ & $1.0$ & $8.5$ & $1.6$ & $6.0$ & -- & -- & -- \\
  	SrTiO$_{3}$$(111)$ & -- & -- & $0.27$ & $1.08$ & $0.33$ & $8.67$ & -- \\
  	\hline \hline
  	\multicolumn{4}{l}{\footnotesize{$^{a}$ From Ref.~\cite{Khalsa2012}}} & 
  	\multicolumn{4}{l}{\footnotesize{$^{b}$ From Ref.~\cite{Santander-Syro2011}}} \\
  	\multicolumn{8}{l}{\footnotesize{$^{c}$ From TB model using experimental masses 
  	                                along $\langle 100 \rangle$}} \\
  \end{tabular}
 \end{center}
\end{table}

For our discussion later, it will be instructive to contrast the observations 
at the $(110)$ and $(111)$ SrTiO$_3$ surfaces with both 
the 2DEG at the $(001)$ surface and a model bulk electronic structure. 
Figure~\ref{fig:fig1}(e) shows the bulk Fermi surface 
from a simplified tight-binding (TB) model 
where the electron hopping amplitudes between the three $t_{2g}$ orbitals 
of neighboring Ti$^{4+}$ are $t_\pi=0.236$~eV and $t_{\delta'}=0.035$~eV~\cite{Khalsa2012},
and we neglect spin-orbit coupling and tetragonal distortions.
Near the $\Gamma$ point, this gives effectives masses listed in the first row 
of table~\ref{Table:Masses} for various directions.
Figures~\ref{fig:fig1}(f,~g) show cross sections of the bulk Fermi surface 
along the $(110)$ and $(111)$ planes through the $\Gamma$ point,
illustrating their respective 2-fold and 6-fold symmetries.
The \emph{experimental} spectra at the SrTiO$_3$ $(001)$ surface~\cite{Santander-Syro2011},
on the other hand,
fit well to a TB form where the hopping amplitudes are
$\bar{t}_\pi=0.36$~eV and $\bar{t}_{\delta'}=0.025$~eV,
leading to values of the effective masses near the $\Gamma$ point
shown in the second rows of table~\ref{Table:Masses}.
Note that all these masses differ by about 30\% from the bulk theoretical ones. 
 
\begin{figure}
  \begin{center}
   	  \includegraphics[clip, width=7cm]{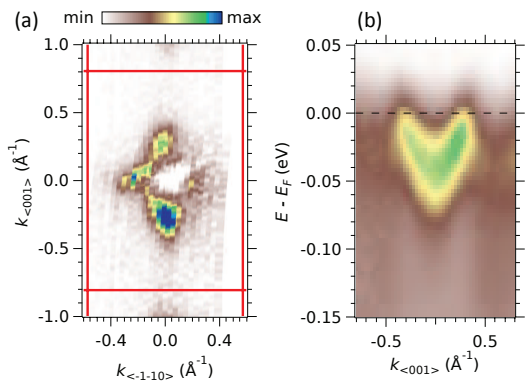}
  \end{center}
  \caption{\label{fig:fig110} \footnotesize{
  		(a)~ARPES Fermi surface map (second derivative) at $h\nu=91$~eV 
  		in the $(110)$ plane of a fractured insulating SrTiO$_3$ sample.
  		The map is a superposition of intensities measured in the bulk $\Gamma_{130}$ 
  		and $\Gamma_{131}$ Brillouin zones~\cite{Supplement}.
  		The red lines indicate the edges of the unreconstructed $(110)$ Brillouin zones.
  		(b)~Energy-momentum intensity map at a $\Gamma$ point 
  		along the $k_{\langle 001 \rangle}$ direction.
  		}
  	} 
\end{figure}

We now present our experimental results.
Figure~\ref{fig:fig110}(a) shows the Fermi surface measured at the \emph{fractured} $(110)$ surface 
of an undoped \emph{insulating} SrTiO$_3$ sample. 
As we will see, our observations are similar 
to another recent study of the 2DEG at the SrTiO$_3$(110) surface in a Nb-doped sample 
prepared \textit{in situ} by Wang~\emph{et al.}~\cite{Wang2013}.
The metallic states we observe present the same 2-fold symmetry of the unreconstructed 
$(110)$ surface Brillouin zone (BZ), represented by red rectangles.
This implies that \emph{(i)} the macroscopic properties of this 2DEG
should be highly anisotropic, echoing the observed anisotropic transport characteristics
reported in 2DEGs at $(110)$ LaAlO$_3$/SrTiO$_3$ interfaces~\cite{Annadi2013},
and \emph{(ii)} any surface roughness or reconstructions, expected in this highly polar surface,
do not affect the 2DEG, which must then reside in the sub-surface layers
--in agreement with our previous conclusions on fractured $(111)$ surfaces of KTaO$_3$~\cite{Bareille2014}. 
Figure~\ref{fig:fig110}(b) shows the dispersion along the $k_{\langle 001 \rangle}$ direction,
giving rise to the longest of the two ellipsoidal Fermi surfaces in figure~\ref{fig:fig110}(a).
The band forming the shortest ellipsoid is eclipsed by photoemission selection rules 
along this direction (see the Supplemental Material~\cite{Supplement}).
The band bottom and Fermi momenta are about $-40$~meV and $0.3$~\AA$^{-1}$, respectively.

From the data above, we model the Fermi surface of the 2DEG at the SrTiO$_3$~$(110)$ surface 
as two orthogonal ellipses, 
one along along $\langle 001 \rangle$ with semi-axes of 0.3~\AA$^{-1}$ and 0.1~\AA$^{-1}$, 
the other along $\langle 1\bar{1}0 \rangle$ with semi-axes 0.25~\AA$^{-1}$ and 0.13~\AA$^{-1}$.
From the area $A_F$ enclosed by the Fermi surfaces, we obtain a carrier density  
$n_{2D}^{(110)} = A_F/2\pi^2 \approx 1 \times 10^{14}$~cm$^{-2}$.
The electronic states associated to such a high charge carrier density 
\emph{must be confined to the region near the surface} 
--otherwise the bulk would be highly conductive, 
in contradiction with the insulating nature of the samples studied.
Similarly, from the band bottom and Fermi momenta, using a parabolic approximation,
we obtain the effective band masses along $\langle 001 \rangle$ 
and $\langle \bar{1}\bar{1}0 \rangle$ (and equivalent directions), 
listed in the third row of table~\ref{Table:Masses}.
These effective masses are similar to the ones determined in the aforementioned study~\cite{Wang2013} 
of the 2DEG at the SrTiO$_3$(110) surface. 
In our study, the band bottom of the heavy band, \emph{c.f.} figure \ref{fig:fig110}(b), 
and the carrier density of the 2DEG are slightly lower, 
probably due to the different surface preparation techniques.

Henceforth, we focus on new experimental results at the $(111)$ surface of SrTiO$_3$, 
which as we will see presents the hexagonal symmetry of the unreconstructed surface, 
and could thus be an interesting platform for the quest
of new electronic states and macroscopic properties at oxide surfaces. 

\begin{figure}
  \begin{center}
   	  \includegraphics[clip, width=7cm]{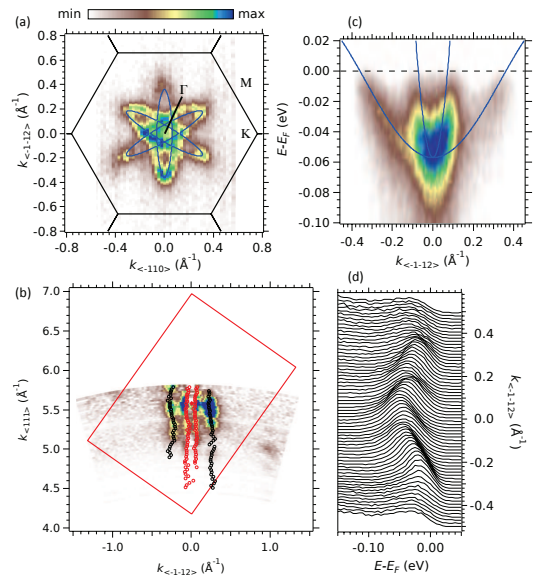}
  \end{center}
  \caption{\label{fig:fig111} \footnotesize{
  		(a) Fermi surface map measured at $h\nu = 110$~eV 
  		on a SrTiO$_3$ $(111)$ surface prepared \emph{in-situ}. 
		The black lines indicate the edges of the unreconstructed $(111)$ 
		Brillouin zones around $\Gamma_{222}$.
  		(b) Fermi surface map (second derivative of ARPES intensity, negative values) 
  		in the $k_{\langle 111 \rangle}$~--~$k_{\langle \bar{1}\bar{1}2 \rangle}$, or $(1\bar{1}0)$ plane, 
  		acquired by measuring at normal emission while varying the photon energy in 1~eV steps 
        between $h\nu_1 = 67$~eV and $h\nu_2 = 120$~eV. 
        The experimental Fermi momenta, represented by the black and red circles, 
        were obtained by fitting the momentum distribution curves (MDCs) 
        integrated over $E_F \pm 5$~meV.
        The red rectangle is the bulk Brillouin zone in the $(1\bar{1}0)$ plane. 
  		(c) Energy-momentum map across the $\Gamma$ point along the $\langle \bar{1}\bar{1}2 \rangle$ direction. 
  		The dispersions of a heavy band and light bands are visible.
  		(d)~Raw energy distribution curves of the dispersions shown in panel~(c).
  		In panels (a) and (c), the blue lines are simultaneous TB fits to the Fermi surface
  		and dispersions.
  		}
  	} 
\end{figure}

Figure~\ref{fig:fig111}(a) shows the Fermi surface measured at the 
SrTiO$_3$~$(111)$ surface prepared \emph{in-situ}, 
as described in the Supplemental Material~\cite{Supplement}.
It consists of three ellipses forming a six-pointed star,
thus strongly differing from the Fermi surface at the SrTiO$_3$~$(110)$ surface,
shown in figure~\ref{fig:fig110}(a), or the one at the SrTiO$_3$~$(001)$ surface,
discussed in previous works~\cite{Santander-Syro2011,Meevasana2011,Plumb2013}.
Additional experiments show that for surfaces prepared \emph{in-situ} with either
$(1\times 1)$ or $(3 \times 3)$ reconstructions,
the band structure and periodicity of the confined states are \emph{identical}, 
and correspond to the one expected from an \emph{unreconstructed surface}~\cite{Supplement}.
This indicates that the 2DEG at the SrTiO$_{3}$$(111)$ surface is also located in the sub-surface layers, 
and is at best weakly affected by the surface reconstructions at the polar $(111)$ surface.

The 2D-like character of the electronic states is strictly demonstrated 
from the Fermi surface map in the 
$\langle 111 \rangle - \langle \bar{1}\bar{1}2 \rangle$ plane, 
shown in figure~\ref{fig:fig111}(b).
Here, one sees that the bands do not disperse along $k_{\langle 111 \rangle}$ 
over more than half a bulk Brillouin zone, 
thereby confirming the confined (\emph{i.e.}, localized) character of the electrons 
along the $[111]$ direction in real space. 
The modulation of the intensity in the Fermi surface map,
a typical feature of quantum well states~\cite{Mugarza2000,Hansen1999}, 
is discussed in the Supplemental Material~\cite{Supplement}.
Interestingly, note that the red rectangles 
in figures~\ref{fig:fig110}(a) and~\ref{fig:fig111}(b) 
represent the Brillouin zone in the $(110)$ (or equivalent) plane. 
Yet, as seen from those figures, the shapes of the corresponding Fermi surfaces 
are completely different. This directly shows the orientational tuning of the Fermi surface 
due to different confinement directions.   

Figure~\ref{fig:fig111}(c) shows the energy-momentum map at the $\Gamma$ point 
along the $\langle \bar{1}\bar{1}2 \rangle$ direction,
corresponding to the major axis of the ellipsoids forming the 6-pointed-star Fermi surface. 
The dispersions of one light band and one heavy band are clearly visible. 
These constitute the ground state of the 2DEG.
Additional subbands are not observed, implying that the band bending at the surface
is too low to populate the upper quantum-well states.
Within our resolution, the heavy and light bands are degenerate at $\Gamma$,
with their band bottom located at about $-57$~meV.
We fit simultaneously these dispersions and the whole Fermi surface of figure~\ref{fig:fig111}(a)
using a simple tight-binding model~\cite{Supplement}.
The fit, shown by the continuous blue lines, 
yields Fermi momenta of about $0.07$~\AA$^{-1}$ and $0.36$~\AA$^{-1}$ 
for, respectively, the light and heavy bands along $\langle \bar{1}\bar{1}2 \rangle$.
This gives an electron concentration 
$n_{2D}^{(111)} \approx 1.0 \times 10^{14}$~cm$^{-2}$,
and effective masses listed in the third row of table~\ref{Table:Masses}.

We now draw some comparisons between the effective masses 
and thicknesses of the 2DEGs at the SrTiO$_3$~$(001)$, $(110)$ and $(111)$ surfaces.
Table~\ref{Table:Masses} shows that, 
while the masses along the ``natural" electron-hopping directions in the bulk 
($[001]$ and equivalent) are comparable between the 2DEGs 
at the SrTiO$_3$~$(001)$ and $(110)$ surfaces, the masses along $[110]$ at the $(110)$ surface,
and all the masses of the 2DEG at the $(111)$ surface,
are very different from the ones expected from the tight-binding parameters 
describing the bulk or the 2DEG at the $(001)$ surface.
In this respect,
note that if the confinement direction is $[110]$ or $[111]$,
then the electrons moving in the 2DEG plane along a direction \emph{other than $[001]$}
will experience the confining potential gradient and the modified crystal field outside the surface,
as they will hop in staircase patterns between first neighbors along $[001]$ 
(or equivalent) directions --see figures~\ref{fig:fig1}(a-d) and Ref.~\cite{Annadi2013,Bareille2014}. 
The understanding of these mass differences,
also reported in quantum well states at thin films of simple-metals~\cite{Wu2002} 
or strongly-correlated oxides~\cite{Yoshimatsu2011}, 
should be the subject of further theoretical works.

The maximal spatial extension $d_{max}$ of the 2DEGs at the SrTiO$_3$ $(110)$ and $(111)$ surfaces
can be estimated using a triangular potential well model~\cite{Supplement}. 
We obtain $d_{max}^{110} \approx 1.7$~nm, which amounts to 6 2D-layers 
or 3 bulk unit cells along $[110]$, 
and $d_{max}^{111} \approx 1.9$~nm, corresponding to $\sim 9$ layers of Ti~$(111)$,
or again about 3 bulk unit cells along $[111]$.

Finally, we note that the orbital ordering of the electronic states 
at the $(110)$ and $(111)$ surfaces of SrTiO$_{3}$ is different from the one at the $(100)$ surface. 
In the first two cases, the bands are degenerate within our experimental resolution, whereas
at the $(001)$ surface the smallest observed splitting between bands 
of different orbital character is of 50~meV~\cite{Santander-Syro2011}.  
As the confinement energy of each band is inversely proportional 
to its effective mass along the confinement direction~\cite{Santander-Syro2011}, 
different surface orientations result in different orbital ordering. 
But along the $[111]$ direction the effective masses of the three $t_{2g}$ bands 
are identical, and so their degeneracy at the $\Gamma$ point is not lifted by the confinement. 
Similarly, the effective masses of bands of different orbital character 
along $[110]$ are quite similar (see table~\ref{Table:Masses}).
Hence, the degeneracy lift is rather small, and cannot be observed in our data. 
This demonstrates the influence of the confinement direction on the orbital ordering.
 
Several scenarios have been proposed to explain the origin of the 2DEG 
at the LaAlO$_3$/SrTiO$_3$ $(001)$ interface. 
According to one of these, the formation of a conducting sheet 
prevents the occurrence of a polar catastrophe in the material. 
Yet, the discovery of a confined 2DEG at the $(001)$ surface of SrTiO$_3$, 
with characteristics similar to those of the above heterostructure, 
suggests that the driving mechanism may not be unique,
as in the bare SrTiO$_3$ all the layers are electrically neutral.
Instead, in the latter case, surface oxygen vacancies are believed to cause 
and to confine the gas~\cite{Santander-Syro2011,Meevasana2011,Wang2013}. 
Additionally, for the $(110)$ and $(111)$ SrTiO$_3$ surfaces, 
of nominal polar charge $4e$, 
one would expect a much larger carrier concentration in the 2DEG, 
and a very strong electric field confining the electrons in a narrow sheet at the surface. 
However, we observe that the carrier concentrations and thicknesses of the 2DEGs
are quite comparable for all three orientations (this work and Ref.~\cite{Santander-Syro2011}):
$n_{2D} \sim 10^{14}$~cm$^{-2}$, $d_{max} \sim 2$~nm.
In fact, in the polar SrTiO$_3$ surfaces studied here, 
the polar catastrophe does not seem to be compensated 
by the electrons of the 2DEG but by surface reconstructions or relaxations,
while the 2DEG lies in the subsurface layers.
Thus, although the 2D electronic structure 
(effective masses, orbital ordering) 
depends on the surface orientation,  
the thickness and carrier concentration of the 2DEG might be controlled by another factor, 
probably oxygen vacancies 
and/or lattice distortions induced 
by the synchrotron light irradiation, as discussed in the Supplemental Material~\cite{Supplement}.

In conclusion, our results show that the symmetries, electronic structure, 
and orbital ordering of the confined states at the surface of TMOs
can be tailored by confining the electrons along different directions 
in the \emph{same} material.
Such orientational tuning echoes the differences of transport properties 
reported recently in LaAlO$_3$/SrTiO$_3$ $(110)$ and $(111)$ 
interfaces~\cite{Herranz2012,Annadi2013,Herranz2013}.
In particular, from our data, the highly anisotropic transport behavior 
observed in the $(110)$ interfaces~\cite{Annadi2013}
can be directly related to the 2-fold symmetry of the Fermi surface measured by ARPES.
More generally, our results provide an exciting route for obtaining 
new types of 2D electronic states in correlated-electron oxides.

We thank V. Pillard for her contribution to the sample preparation.
T.C.R. acknowledges funding from the RTRA Triangle de la Physique (project PEGASOS).  
A.F.S.-S. and M.G. acknowledge support from the Institut Universitaire de France.
This work is supported by public grants from the French National Research Agency (ANR) 
(project LACUNES No ANR-13-BS04-0006-01) 
and the ``Laboratoire d'Excellence Physique Atomes Lumi\`ere Mati\`ere'' (LabEx PALM project ELECTROX) 
overseen by the ANR as part of the ``Investissements d'Avenir'' program (reference: ANR-10-LABX-0039).

\section{Supplemental Material}

\subsection*{ARPES Experiments}
The ARPES measurements were conducted at the Synchrotron Radiation Center 
(SRC, University of Wisconsin, Madison) 
and the Synchrotron Soleil (France). 
We used linearly polarized photons in the energy range $20-120$~eV, 
and Scienta R4000 electron detectors with vertical slits. 
The angle and energy resolutions were $0.25^{\circ}$ and 25~meV at SRC,
and $0.25^{\circ}$ and 15~meV at Soleil.
The mean diameter of the incident photon beam was smaller than 100~$\mu$m.
The samples were cooled down to 10-30~K before fracturing or measuring, 
in pressure lower than $6\times10^{-11}$~Torr. 
The confined states were either created by fracturing the samples in vacuum 
or by chemically and thermally preparing the surfaces \emph{in situ},
as detailed in the next section.
The results were reproduced for at least five different samples for each surface orientation. 

\subsection*{Surface preparation}
\begin{figure*}
  \begin{center}
   	  \includegraphics[clip, width=14cm]{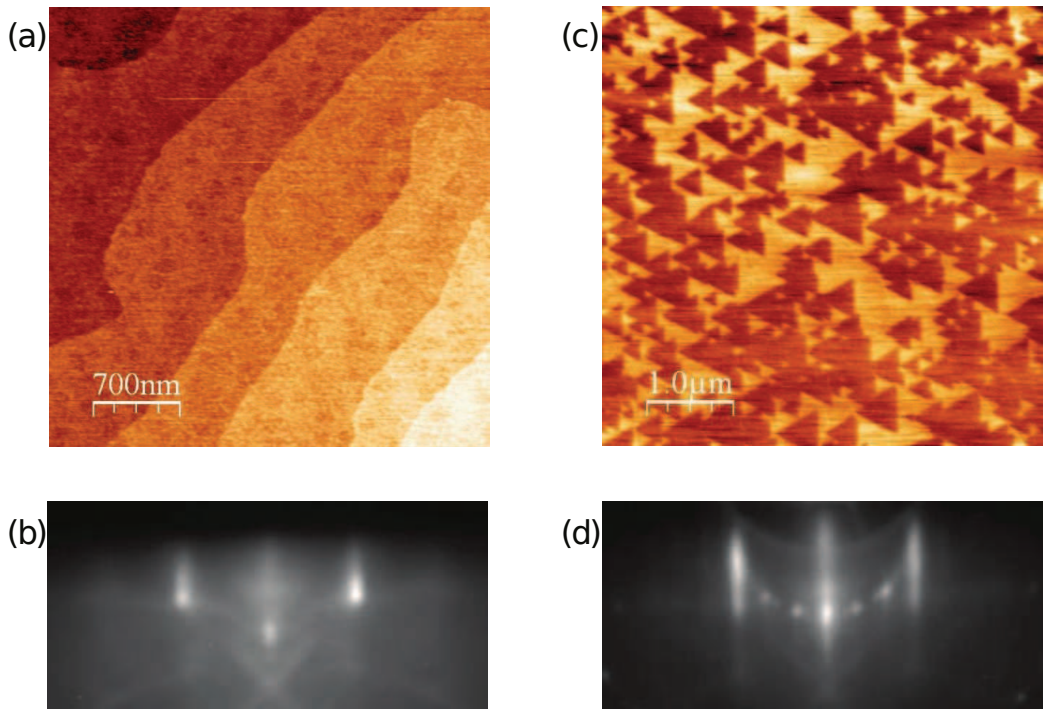}
  \end{center}
  \caption{\label{fig:sto111_afm} \footnotesize{(Color online) 
  		(a) Atomic force microscope (AFM) image of a chemically 
  		and thermally prepared SrTiO$_3$~$(111)$ surface. 
  		The surface is single terminated and unreconstructed, 
  		as shown in the RHEED image in (b). 
        Longer annealing times result in a mixed terminated surface, 
        as demonstrated in the AFM friction image (c) measured in contact mode. 
        A $3 \times 3$ reconstruction of the surface can be deduced 
        from the corresponding RHEED image in (d).
  		}
  	} 
\end{figure*}

The non-doped, polished crystals of SrTiO$_3$ were supplied by CrysTec GmbH and Aldrich.
To prepare the surface, the samples were ultrasonically agitated in deionized water, 
subsequently etched in buffered HF and annealed at $950^{\circ}$C 
for several hours in oxygen flow. 
Depending on the annealing time, this treatment yields a Ti-rich, 
single-terminated or mixed-terminated step-and-terrace structured 
surface of SrTiO$_3$~$(111)$~\cite{Biswas2011}. 
Figure~\ref{fig:sto111_afm}(a) shows the atomic-force microscopy (AFM) image 
of the single-terminated $(111)$ surface of a sample annealed for 3h. 
This treatment produces a $(1 \times 1)$ unreconstructed surface, 
shown by RHEED image in figure~\ref{fig:sto111_afm}(b).
Longer annealing (10h) results in a mixed-terminated surface~\cite{Chang2008},
as shown in the AFM friction image in figure~\ref{fig:sto111_afm}(c), 
measured in contact mode. 
The surface prepared in such a way is $(3 \times 3)$ reconstructed, 
as displayed in the RHEED image in figure~\ref{fig:sto111_afm}(d). 
The surface state of the cleaved samples 
was not determined by imaging or diffraction techniques.

To perform the surface-sensitive ARPES measurements, 
one needs pristine and crystalline surfaces. 
To clean the surface of contaminations, the samples prepared as described above 
were further annealed \emph{in-situ} in vacuum at a pressure of 
approximately $p=3\times10^{-9}$~mbar at a temperature of $T = 550^\circ$C for about 2 hours. 
This annealing step cleans the surface, does not change the surface reconstruction, 
and also introduces oxygen vacancies in the bulk of the SrTiO$_3$ samples. 
Note that the introduced bulk charge carrier density is at least 
three orders of magnitude lower than the one observed for the confined states 
in the ARPES measurements, as detailed in the main text.  
Moreover, Plumb~\emph{et al.} demonstrated that various 
\emph{in-situ} sample preparations, including annealing in an O$_2$-rich atmosphere 
which results in a non-doped bulk, create identical confined states 
at the $(001)$ surface of TiO$_2$-terminated SrTiO$_3$ ~\cite{Plumb2013}.
Recall also, from figure~\ref{fig:fig111}(b), that the states observed in our experiments 
do not disperse along the confinement direction, which demonstrates
their quasi-2D character.

For the confined states at the $(111)$ surface, the quality of the obtained ARPES data 
is better for the surface prepared \emph{in-situ}. 
This might be due to the strong polar nature of the $(111)$ surface of SrTiO$_3$. 
Hence, fracturing a sample along a $(111)$ plane might yield a partly disordered surface.

The electronic structure of the 2DEG at the SrTiO$_3$~$(111)$ surface
is similar for the cleaved and the two differently prepared surfaces 
(unreconstructed and $(3 \times 3)$ reconstructed). 
In fact, for all three types of surfaces the periodicity of the electronic structure 
in reciprocal space, shown in figure~\ref{fig:fig_sup}(a) for the prepared, $(3 \times 3)$
reconstructed surface, corresponds to the one expected of an unreconstructed surface.
By Bloch theorem, the very existence of dispersive bands and well-defined Fermi surfaces 
implies the existence of a periodic in-plane potential acting on the confined electrons,
hence of crystalline order at the layer(s) where the 2DEG is located.
As the electronic structure has the periodicity of the \emph{unreconstructed} surface,
the 2DEG seems to stabilize in a sub-surface region, where it is not affected by
any surface reconstructions or superstructures related to vicinal surfaces or terraces.
A possible explanation for this observation would be that the electrons of the Ti cations 
in the topmost layer are localized, while the itinerant electrons exist in the subsurface layers.
For the $(110)$ surface, a surface preparation similar to the one described above 
for the $(111)$ surface was conducted.  
The data quality of fractured and prepared samples are quite similar 
as the chemical etching step is not perfectly adapted to the $(110)$ surface. 
Sr and Ti are both situated in one of the alternating $(110)$ layers 
of $($SrTiO$)^{4+}$ and O$_2^{4-}$ building up the crystal lattice. 
Thus, the selective etching of Sr-related species might result in a rather rough surface.

\subsection*{Photon energy dependence}
\begin{figure}
    \begin{center}
        \includegraphics[clip, width=7cm]{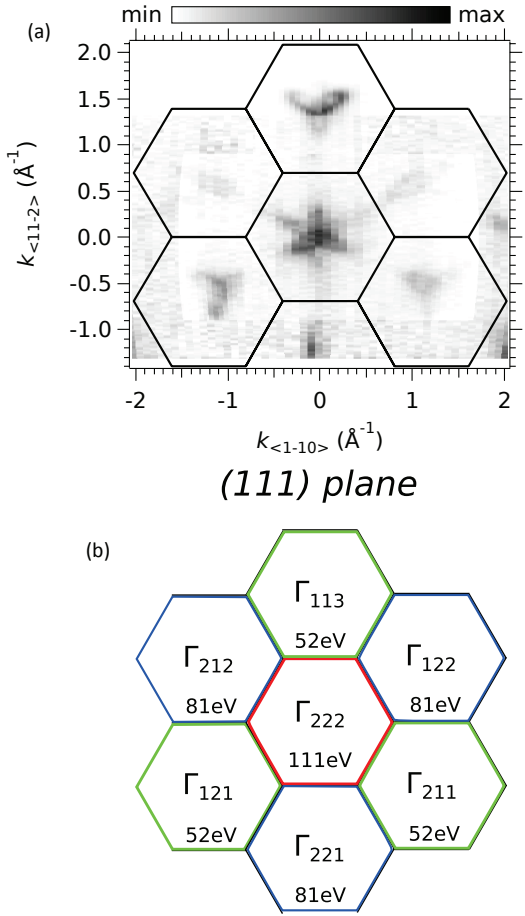}
    \end{center}
    \caption{\label{fig:fig_sup} \footnotesize{(Color online) 
        (a) Superposition of Fermi surface maps measured 
        for the chemically and thermally prepared SrTiO$_3$~$(111)$ sample 
        ($(3\times 3)$ reconstructed surface) 
        at photon energies of $h\nu = 47$~eV and $h\nu=96$~eV. 
        (b) Reciprocal 2D space in the $(111)$ plane. 
        Inside each Brillouin zone the projections of the different bulk $\Gamma$ points 
        corresponding to available final states during the photoemission
        process at the specified photon energy are indicated. 
        This diagram helps understanding the Fermi-surface intensities shown in panel (a).
        The color (red, blue, green) of the hexagons 
        indicates which $\Gamma$ points are located in the same (111) plane in reciprocal space.
        }
      }     
\end{figure}

The photon energy dependence of the electronic states at the SrTiO$_3$~$(111)$ surface 
is displayed in the main text in figure~\ref{fig:fig111}. 
Although the states do not disperse, confirming their confined nature, 
the intensity of the states drops rather quickly moving away from $\Gamma_{222}$.
This observation is similar to the intensity modulation as a function of the photon energy 
reported previously at the $(001)$ surface of SrTiO$_3$~\cite{Santander-Syro2011} 
and KTaO$_3$~\cite{Santander-Syro2012}, as well as in quantum well states 
of metals~\cite{Mugarza2000,Hansen1999}. 
This modulation is due to photoemission dipole selection rules: 
the optical excitation of the electrons occurs from initial states 
in the near surface region that do not disperse along the confinement direction 
(the confined electrons) to dispersing bulk final states.
Moreover, if the wave function of the confined states is not exactly 
localized in a 2D layer, but exists over several unit cells, 
the dispersion along the confinement direction will be affected.
This can be intuitively understood from Heisenberg uncertainty principle:
only a strict 2D confinement in real space yields a complete indetermination
of the electron momentum along the confinement direction, 
hence an exactly cylindrical Fermi surface.
Some delocalization along the confinement direction, as in quantum-well states, 
implies a small dispersion
of the Fermi surface along that direction. 
    
Bearing these effects (selection rules in quantum wells, finite delocalization) in mind, 
one can comprehend the data in figure~\ref{fig:fig_sup}(a),
which shows a superposition of Fermi surface maps measured at different photon energies, 
for a $(111)$ surface prepared \emph{in-situ}. 
The black hexagons are the Brillouin zones 
assuming an unreconstructed surface. 
Thus, due to selection rules, the intensity of the photoemission peak 
from the confined states is highest close to positions 
corresponding to $\Gamma$ points of the bulk,
where final states at the same $k_{\langle 111 \rangle}$ momentum are available for the optical transition. 
But this intensity will decrease rapidly by moving along $k_{\langle 111 \rangle}$, 
away from the bulk $\Gamma$ points~\cite{Mugarza2000}. 
Experimentally, this is done by changing the photon energy. 
This results in the necessity to measure in-plane Fermi surface maps 
at different photon energies, and then superpose them
to retrieve the complete periodicity of the electronic states,
as illustrated in figure~\ref{fig:fig_sup}(b). This figure shows the positions 
of the experimentally observed $\Gamma$ points projected in the $(111)$ plane. 
The photon energy inside each Brillouin zone 
corresponds to the $k_{\bot}$ value of the $\Gamma$ points 
assuming a work function of $W = 4.25$~eV and an inner potential of $V_{0} = 12$~eV. 

\subsection*{Fermi surface of S\lowercase{r}T\lowercase{i}O$_3$(110)}
\begin{figure}
  \begin{center}
   	  \includegraphics[clip, width=5cm]{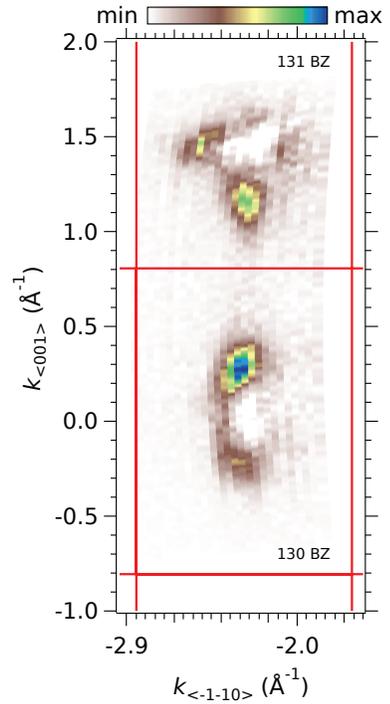}
  \end{center}
  \caption{\label{fig:sto110_unfoldedFS} \footnotesize{(Color online) 
  		Second derivative of ARPES Fermi surface map at $h\nu=91$~eV 
  		in the $(110)$ plane of a cleaved insulating SrTiO$_3$ sample.
  		The map spans the $\Gamma_{130}$ (bottom)
  		and $\Gamma_{131}$ (top) Brillouin zones.
  		The red lines indicate the edges of the unreconstructed $(110)$ Brillouin zones.
  		}
  	} 
\end{figure}
As stated in the main text, 
the Fermi surface map shown in figure~\ref{fig:fig110}
is a superposition of intensities measured in the bulk $\Gamma_{130}$ and $\Gamma_{131}$ Brillouin zones. 
Figure~\ref{fig:sto110_unfoldedFS} shows the intensities measured 
in those Brillouin zones.
Due to photoemission matrix elements, only the vertical ellipsoidal Fermi surface 
is observed around the $\Gamma_{130}$ point, 
while both the vertical and the smaller horizontal ellipsoidal Fermi surfaces 
are observed around the $\Gamma_{131}$ point.

\subsection*{Estimate of the spatial extensions of the 2DEGs at the S\lowercase{r}T\lowercase{i}O$_3$~$(110)$ and $(111)$ surfaces.} 
In our data, figures~2 and~3 of the main text, only the lowest-energy subbands are observed.
To estimate the maximal extension $d_{max}$ of the corresponding confined states, 
we follow the same strategy of Ref.~\cite{Bareille2014}.
We assume that the second subbands are slightly above the Fermi level, 
hence unoccupied and not detectable by ARPES. 
We then use a triangular potential well model, and take as effective masses
along the $[110]$ and $[111]$ confinement directions, respectively,
$m_{110} \approx 1.6 m_e$ 
(the lightest of the masses gives the largest 2DEG thickness) 
and $m_{111} = 1.0 m_e$ (given by extrapolating the experimental masses at the $(001)$ surface
to the bulk $[111]$ direction) --see table~I of the main text.
This gives $d_{max}^{110} \approx 1.7$~nm, amounting to 6 2D-layers 
or 3 bulk unit cells along $[110]$, 
and $d_{max}^{111} \approx 1.9$~nm, corresponding to $\sim 9$ layers of Ti~$(111)$,
or again about 3 bulk unit cells along $[111]$.

\subsection*{UV dose dependence: enhancement of T\lowercase{i}$^{3+}$ signal} 
\begin{figure}
    \begin{center}
        \includegraphics[clip, width=8cm]{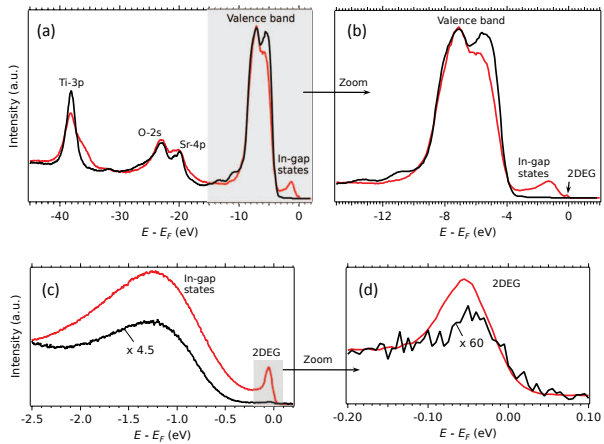}
    \end{center}
    \caption{\label{fig:fig_UVdose} \footnotesize{(Color online) 
        (a) Angle-integrated spectra of an SrTiO$_3$ sample prepared \emph{in-situ}, 
        measured at a photon energy of $hv=110$~eV, with a step size of 50~meV,
        showing the density of states for binding energies between $-45$~eV and $2$~eV.
        The black curve was measured shortly after the first exposure of the sample 
        to the UV light, and the red curve at the end of the measurements (about 36 hours later). 
        (b) Zoom over the valence band region. 
        (c) Angle-integrated spectra showing the in-gap states 
        and the confined states at the Fermi level, 
        measured at $hv=110$~eV with a step size of 5meV. 
        (d) Zoom over the confined states at the Fermi level.
        }
      }     
\end{figure}

Understanding the influence of the UV synchrotron illumination 
on the observed confined states is important to determine the origin of such states. 
Recent photoemission studies on the 2DEGs 
at the $(001)$ or $(110)$ surface of SrTiO$_3$ 
proposed that the UV light creates oxygen vacancies~\cite{Meevasana2011,Wang2013} or, respectively, 
ferroelectric lattice distortions~\cite{Plumb2013} in the surface region. 
The two effects are difficult to disentangle using photoemission, 
as in both cases charge is transferred from O to Ti.
Figure\ref{fig:fig_UVdose}(a) shows the angle-integrated spectra, 
measured at $hv=110$~eV, of a SrTiO$_3$ sample prepared \emph{in-situ}
for binding energies between $-45$~eV and $2$~eV.
The black curve was measured 
shortly after the first exposure of the sample to the UV light,
while the red curve was recorded at the end of the measurements (36 hours later). 
The spectra are normalized to the intensity of the Sr $4p$-peak, 
which should be rather independent of the concentration of oxygen vacancies and/or 
ferroelectric lattice distortions. 
Figure~\ref{fig:fig_UVdose}(b) is a zoom over the valence band region, 
while figures~\ref{fig:fig_UVdose}(c,~d) show the in-gap states and the 
confined states at the Fermi level. 
The change of various features under UV irradiation is obvious: 
first, the formation of a shoulder in the Ti-$3p$ peak 
at lower binding energies, indicating electron transfer from Ti$^{4+}$ to a lower valency state. 
Second, the decrease in intensity of the valence band in its low binding energy region. 
Third, the increase in intensity of the in-gap states 
and of the peak corresponding to the confined states. 
All these observations could be explained by both scenarios: 
the creation of oxygen vacancies and the ferroelectric lattice distortions. 

In contrast to samples prepared \emph{in-situ}, 
cleaved samples show a different behavior regarding the UV light exposure. 
The subbands of the 2DEG in all the \emph{cleaved} SrTiO$_3$ surfaces 
we have studied so far, \emph{i.e.} $(001)$, $(110)$ and $(111)$, 
are all observed essentially \emph{immediately after cleaving},
with no or little time delay after the first exposure to UV light. 
A more detailed study on the UV induced effects is beyond the scope of this paper.      

\subsection*{Tight-binding calculations of the 2DEG at the S\lowercase{r}T\lowercase{i}O$_3$$(111)$ surface} 
The band dispersions shown in the main text correspond 
to the bottom of the conduction band of SrTiO$_3$,
which is formed by Ti-$3d$ orbitals hybridized with O-$2p$ orbitals. 
The interaction between the oxygen anions forming an octahedron 
and the Sr cation generates a large crystal field which splits the $d$ states 
in a lower $t_{2g}$ triplet and an higher $e_{g}$ doublet. 
Hence, only the $t_{2g}$ orbitals are considered in our tight-binding model, 
which is based on the calculations of reference~\cite{Xiao2011}. 
Our model for the SrTiO$_3$$(111)$ surface is limited to a bilayer of Ti atoms. 
This approach is sufficient to fit the experimental data as shown in the main text, 
but does not necessarily imply the confinement of the electrons to a bilayer.

The Hamiltonian $H$ of the system in the basis $\{d_{I,n}\}$, 
where $I=(X,Y,Z)$ correspond to the orbital character $(yz,zx,xy)$ of the $t_{2g}$ orbitals 
and $n=1,2$ indicates the number of the layer of Ti cations, is given by:

\begin{equation*}
H=
    \begin{pmatrix}
        d^\dagger_{X,1} \\
        d^\dagger_{Y,1} \\
        d^\dagger_{Z,1} \\
        d^\dagger_{X,2} \\
        d^\dagger_{Y,2} \\
        d^\dagger_{Z,2} 
    \end{pmatrix}
    ^T
    \begin{pmatrix}
        \tilde{\epsilon}_X &  &  &  \epsilon_{X} & & \\
        & \tilde{\epsilon}_Y & & &  \epsilon_{Y} & \\
        & & \tilde{\epsilon}_Z & & &  \epsilon_{Z}  \\ 
        \epsilon_X^* &  &  &  \tilde{\epsilon}_{X} & & \\
        &  \epsilon_Y^* & & &  \tilde{\epsilon}_{Y} & \\
        & &  \epsilon_Z^* & & &  \tilde{\epsilon}_{Z}  \\
    \end{pmatrix}
    \begin{pmatrix}
        d_{X,1} \\
        d_{Y,1} \\
        d_{Z,1} \\
        d_{X,2} \\
        d_{Y,2} \\
        d_{Z,2}. 
    \end{pmatrix}
\end{equation*}

Here, $\tilde{\epsilon}_I$ describes the hopping of electrons 
between next nearest neighbors of Ti cations (intra-layer hopping), 
characterized by the hopping amplitude $t_{\sigma''}$, 
whereas $\epsilon_I$ describes the hopping between nearest neighbors 
(inter-layer hopping) with hopping amplitudes $t_\pi$ and $t{_\delta'}$:

\begin{align*}
    \tilde{\epsilon}_X&=-2t_{\sigma''}\cos(-\frac{\sqrt{3}}{2}\tilde{a}~k_x + \frac{3}{2}b~k_y)\\
    \tilde{\epsilon}_Y&=-2t_{\sigma''}\cos(\frac{\sqrt{3}}{2}\tilde{a}~k_x + \frac{3}{2}b~k_y)\\
    \tilde{\epsilon}_Z&=-2t_{\sigma''}\cos(\sqrt{3}\tilde{a}~k_x)\\
    \epsilon_X&=-t_\pi e^{-i\tilde{a}~k_y}
        \left[1+e^{i\frac{\tilde{a}}{2}(-\sqrt{3}k_x+3k_y)}\right]
        -t_{\delta'}e^{i\frac{\tilde{a}}{2}(\sqrt{3}k_x+k_y)}\\
    \epsilon_Y&=-t_\pi e^{-i\tilde{a}~k_y}
        \left[1+e^{i\frac{\tilde{a}}{2}(\sqrt{3}k_x+3k_y)}\right]
        -t_{\delta'}e^{i\frac{\tilde{a}}{2}(-\sqrt{3}k_x+k_y)}\\
    \epsilon_Z&=-2t_\pi e^{\frac{i}{2}\tilde{a}~k_y}
        \cos(\frac{\sqrt{3}}{2}\tilde{a}~k_x)-t_{\delta'} e^{-i\tilde{a}~k_y}.
\end{align*}

In the above expressions, $k_x$ corresponds to $k_{\langle 1\bar{1}0 \rangle}$, 
$k_y$ to $k_{\bar{1}\bar{1}2}$, and $\tilde{a}$ to the cubic lattice constant $a$ 
projected in the (111) plane $\tilde{a}=\sqrt{2/3}a$. 
Compared to the calculations of reference~\cite{Xiao2011}, our data can be fitted rather well 
using a simplified model. We neglect in our model the spin-orbit coupling, 
the trigonal crystal field, the layer potential difference, 
crystal distortions at low temperature, and the hopping ($t_{\pi'}$) 
between next nearest neighbors of different orbital symmetry. 
The fits shown in figures~\ref{fig:fig111}(a) and \ref{fig:fig111}(c) of the main text 
are based on such a simplified model using fitting parameters of 
$t_\pi=1.6$~eV, $t_{\delta'}=0.07$~eV and $t_{\sigma''}=0.05$~eV.

Note that such value of $t_\pi$, which quantifies the hopping energy between nearest neighbors
along the $[100]$ (and equivalent) directions, is here over 4 times larger than
the same parameter inferred from the 2DEG at the SrTiO$_3$~$(001)$ surface 
(namely, $t=0.36$~eV, see the main text).  This shows again that the effective masses
of the 2DEG at the SrTiO$_3$~$(111)$ surface strongly differ from what would be
expected from a model based on the 2DEG at the $(001)$ surface.
As discussed in the main text, the electrons moving along any direction in the $(111)$ plane
will actually hop in zig-zag patterns between first neighbors along $[001]$ 
(or equivalent) directions, and thus will experience the confining potential gradient 
and the modified crystal field outside the surface.  These effects are not accounted by
our minimalist TB model.
Additionally, our TB model only considers one bilayer of Ti atoms. 
However, it is known that in quantum well states the effective masses of the confined electrons 
depend on the width of the quantum well or, equivalently, 
the number of layers~\cite{Wu2002,Yoshimatsu2011}.
All these effects should be taken into account in future theoretical works addressing the 2DEGs at
the different surfaces of SrTiO$_3$. 
On the other hand, while distortions of the crystal lattice, 
and thereby of the overlap between the different $t_{2g}$ orbitals, 
might exist at the surface and be slightly different depending on the surface orientations, 
they should bear a negligible effect on the 2DEGs reported here, as we have seen that
their electronic structure is essentially insensitive to surface polarity or reconstructions. 



\end{document}